\documentclass{pasj00}
\draft

\begin{document}
\SetRunningHead{Nakajima and Tsuru et al. }{X-Ray Reflection Nebulae 
  in the Sgr C Region}
\Received{XXXX/XX/XX}%{yyyy/mm/dd}
\Accepted{YYYY/YY/YY}%{yyyy/mm/dd}

% \title{Discoveries of Diffuse Structures from the Sgr C Region}
\title{X-Ray Reflection Nebulae 
  with Large Equivalent Widths of Neutral Iron K$\alpha$ Line
  in the Sgr C Region}

%%% begin:list of authors
% Do NOT capitalize all letters in "textsc".
%
\author{%
  Hiroshi \textsc{Nakajima}}
\affil{%
  Department of Earth and Space Science, Graduate School of Science, 
  Osaka University, Toyonaka, Osaka, 560-0043}
\email{nakajima@ess.sci.osaka-u.ac.jp}
\author{%
  Takeshi Go \textsc{Tsuru},
  Masayoshi \textsc{Nobukawa},
  Hironori \textsc{Matsumoto},
\and
  Katsuji \textsc{Koyama}}
\affil{%
   Department of Physics, Graduate School of Science, Kyoto University,
   Sakyo-ku, Kyoto, 606-8502}
\author{%  
  Hiroshi \textsc{Murakami}}
\affil{%
   Department of Physics, Rikkyo University, 3-34-1, Nishi-Ikebukuro,
   Toshima-ku, Tokyo, 171-8501}
\author{%
  Atsushi \textsc{Senda}}
\affil{%
   RIKEN (The Institute of Physical and Chemical Research), 2-1,
   Hirosawa, Wako, Saitama, 351-0198}
\and
\author{%
  Shigeo \textsc{Yamauchi}}
\affil{%
   Faculty of Humanities and Social Sciences, Iwate University, 3-18-34,
   Ueda, Morioka, Iwate, 020-8550}

%% `\KeyWords{}' always has to be placed before `\maketitle'.
%Do NOT move this preamble from here!

\maketitle

\begin{abstract}
This paper reports on the first results of the Suzaku observation 
in the Sgr~C region. We detected four diffuse clumps with strong
line emission at 6.4~keV, K$\alpha$ from neutral or low-ionized Fe.
One of them, M\,359.38$-$0.00, is newly discovered with Suzaku.
The X-ray spectra of the two bright clumps, M\,359.43$-$0.07 and M\,359.47$-$0.15,
after subtracting the Galactic center diffuse X-ray emission (GCDX),
exhibit strong K$\alpha$ line from Fe\emissiontype{I} with large
equivalent widths (EWs) of 2.0--2.2~keV and clear K$\beta$ of Fe\emissiontype{I}.
The GCDX in the Sgr~C region is composed of the 6.4~keV- and 6.7 keV-associated
components. These are phenomenologically decomposed by taking relations
between EWs of the 6.4~keV and 6.7~keV lines. Then the former EWs against the associated
continuum in the bright clump regions are estimated to be $2.4^{+2.3}_{-0.7}$~keV.
Since the two different approaches give similar large EWs of 2~keV,
we strongly suggest that the 6.4~keV clumps in the Sgr~C region are due to X-ray 
reflection/fluorescence (the X-ray reflection nebulae). 
\end{abstract}

% Introduction
%
\section{Introduction}

%GCの活動牲
The Galactic center (GC), the nearest galactic nucleus, hosts various classes of
high energy sources such as super massive black hole Sgr~A$^*$, X-ray binaries,
supernova remnants, and non-thermal filaments (NTFs).
GC is also the primary site of giant molecular 
clouds (MCs) and star formation inside the MCs. 
One of the most interesting characteristics in GC is the diffuse 6.4~keV emission 
from neutral iron (Fe\emissiontype{I}). 
Most of the prominent 6.4~keV emissions are found in the 
MCs of Sgr~B2 \citep{Koyama96}, G\,0.13$-$0.13 \citep{Yusef02},
Sgr~B1 \citep{Nobukawa08}, and  Sgr~C \citep{Murakami01b}. 
A number of small MCs in the GC region also emit strong 6.4~keV line
\citep{Predehl03,Park04}.

%6.4keV clumpとその起源
For the origin of 6.4~keV emission line from MCs,
two models have been proposed.
One is the impact of low-energy cosmic-ray electrons (LECRe)
followed by bremsstrahlung and 6.4~keV line emission,
proposed for the origin of the Galactic Ridge X-ray Emission \citep{Valinia00}.
\citet{Yusef02} proved that this model can be applied to G\,0.13$-$0.13
assuming the Fe abundance of two solar.
The other is that the hard X-rays and 6.4~keV lines are due to
reflection/fluorescence from  MCs irradiated by external hard X-ray sources
(X-ray reflection Nebulae; XRNe) \citep{Sunyaev93,Markevitch93,Koyama96,Sunyaev98,Park04}.
Since no irradiating source capable of powering the 6.4~keV line was found,
\citet{Koyama96} proposed a scenario of
a past X-ray outburst of the super massive black hole at Sgr~A$^*$.
The present luminosity is $2\times 10^{33}$erg~s$^{-1}$
\citep{Baganoff01, Baganoff03}, and hence it would have been $10^6$ times
brighter than now 300 years ago.

In fact, the Chandra observation shows that the 6.4~keV emission line of the giant
MC Sgr~B2 has a concave shape pointing to the GC direction \citep{Murakami01a}.
The broad band spectrum of Sgr~B2 is well described by Compton scattered and
reprocessed radiation emitted in the past by Sgr~A$^*$ \citep{Revnivtsev04}.
The morphology and flux of the 6.40~keV line of Sgr~B2
have been time variable for 10~years \citep{Muno07,Koyama08a, Inui08}.
Thus observational results supporting the XRN
rather than the LECRe model have been accumulating for Sgr~B2.
However, for the other regions, in particular Sgr~C, observations
have been limited and hence the 6.4~keV line origin is still open issue.

%過去のSgrC X線観測とすざくでの観測のモチベーション
The Sgr~C region consists of giant MCs and large H\emissiontype{II} regions.
From the $^{13}$CO observations, \citet{Liszt95} estimated
$6.1\times 10^{5}$ solar mass for the molecular gas.
\citet{Murakami01b} discovered a 6.4 keV clump with ASCA, and
interpreted that the clump is an XRN because
the X-ray spectrum exhibited a large equivalent width (EW) of the 6.4~keV
emission line with a strong absorption like that in Sgr~B2.
On the other hand, \citet{Yusef07} discussed possible association of
the 6.4~keV emission lines with some radio NTFs based on the Chandra observation,
and argued that the 6.4~keV line is due to the impact of LECRe.
These results and arguments are based on the short exposure
observations (20~ksec and 22~ksec for the ASCA and Chandra, respectively).
In order to fix the above debates, we, therefore,
made a long Suzaku observation on the Sgr~C region.

For simplicity,  we use ``the east'' as the positive Galactic longitude side,
and "the north" as the positive Galactic latitude side.
We assume the distance to the Sgr~C region to be 8.5~kpc
following the recommendation of IAU.
We applied the solar abundance and photoelectric absorption
cross-section from the tables given by \citet{Anders89} and \citet{Church92}, respectively.

%%%%%%%%%%%%%%%%%%%%%%%%%%%%%%%%%%%%%
% Observation
%
\section{Observation and Data Reduction}\label{ssec:suzakuobs}
The Sgr~C region was observed with the XIS (X-ray Imaging Spectrometer) 
from 2006 Feb. 20 to 23.
The telescope optical axis position was R.A.=\timeform{17h44m37s.30},
Decl.= \timeform{-29D28'10''.2} (J2000.0).
The XIS consists of four sets of the X-ray CCD camera systems placed on the 
focal planes of the four X-Ray Telescopes (XRT) onboard the Suzaku satellite.
One of the XIS sensors (XIS1) has a back-illuminated (BI) CCD, while 
the other three sensors (XIS0, 2 and 3) utilize front-illuminated (FI) CCDs. 
The detailed descriptions of the Suzaku satellite, XRT, and XIS can be found in
\citet{Mitsuda07}, \citet{Serlemitsos07}, and \citet{Koyama07a}, respectively.

%データプロセスとスクリーニング
A cleaned event list was obtained from the processed data with the
version of 2.0.6.13\footnote{http://www.astro.isas.jaxa.jp/suzaku/process/}
by removing events taken during the passage of the South Atlantic Anomaly,
the elevation angles from the night Earth rim of $<\timeform{5D}$ 
and from the sun-lit Earth rim of $<\timeform{20D}$, and the telemetry saturation. 
After these data screenings, the net exposure was 107~ksec. 
We analyzed the screened data using the HEADAS software version 6.4, XSPEC
version 11.3.2g\footnote{http://heasarc.gsfc.nasa.gov/docs/software/lheasoft/}.
We utilized the calibration databases released on 2008 Feb. 1
\footnote{http://www.astro.isas.jaxa.jp/suzaku/caldb/}. 

%%%%%%%%%%%%%%%%%%%%%%%%%%%%%%%%%%%%%
% Analysis
%
\section{Analysis and Results}

The non X-ray (particle) background (NXB) becomes serious
in the hard X-ray band, in particular for diffuse sources.
The flux of the NXB is correlated with the geomagnetic cut-off rigidity (COR)
\citep{Tawa08}. We, therefore, made COR-sorted NXB data sets using the night-Earth
data released by the Suzaku XIS
team \footnote{http://www.astro.isas.jaxa.jp/suzaku/analysis/xis/nte/}. 
For the following imaging and spectral studies in the Sgr~C region
(section~\ref{ssec:image} and \ref{ssec:specanaI}), 
we used the X-ray data after subtracting the NXB that were compiled
to have the same COR distribution as that during the Sgr~C observation.

To increase statistics, hear and after, we co-added the four XIS data for imaging study,
in which exposure and vignetting correction were made.
For the spectral study, we co-added the three FI CCD data, and treated BI data
separately
because the response functions are significantly different between FI and BI.

\subsection{X-ray Image in the Emission Line}\label{ssec:image}

\begin{figure}
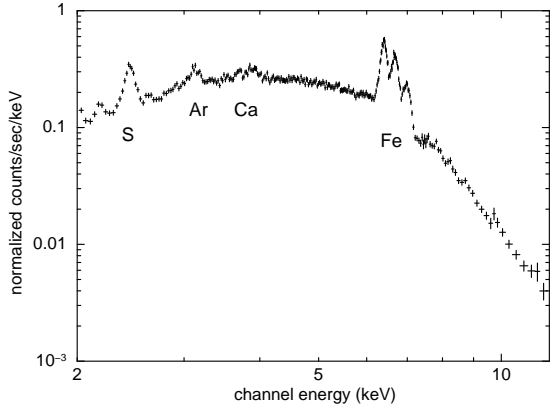

  \begin{center}
%   \rotatebox{-90}{\FigureFile(54mm,45mm){SgrCentirefit_080630.ps}}
   \rotatebox{-90}{\FigureFile(54mm,45mm){figure1.ps}}
  \end{center}
  \caption{The co-added GCDX spectrum of the FIs (XIS0+2+3) 
    in the Sgr~C region extracted from the \timeform{6'.95} radius 
    circle in the center of the XIS FOV. 
    The NXB component was subtracted. 
    The spectrum of BI (XIS1)  was simultaneously analyzed,
    but the figure is not shown here, for brevity.
    }
  \label{fig:Spec_circle400_xis023_grp500}
\end{figure}

Figure~\ref{fig:Spec_circle400_xis023_grp500} shows the energy spectrum 
of the whole Sgr~C region, which was extracted from the circular region 
with a radius of $\timeform{6'.95}$ from the FOV center. 
The spectrum shows K-shell emission lines from He-like and/or H-like ions of 
S, Ar, Ca, and Fe.
In addition, the 6.4~keV emission line and the K-edge absorption 
at 7.1~keV due to Fe\emissiontype{I} are detected. 
The three Fe K-shell emission lines at 6.4, 6.7 and 6.9~keV in the 
Galactic center diffuse X-ray emission (GCDX)\citep{Koyama89, Yamauchi90, Koyama07b}
are clearly resolved.

Since the Suzaku coordinate has 90\% confident
error of \timeform{19''} \citep{Uchiyama08},
we made coordinate correction using a Chandra observation.
In the energy band of 0.7--1.5~keV, a point source
with the flux of $\sim~6\times~10^{-15}$~erg~cm$^{-2}$~s$^{-1}$ was detected
at (R.~$\!\!$A., Decl.)$_{\rm{J2000.0}}$=(\timeform{17h44m52s.1}, \timeform{-29D31'56''.7})
in the XIS image. In the error circle of Suzaku, there was only one Chandra point
source at (R.~$\!\!$A.,Decl.)$_{\rm{J2000.0}}$=(\timeform{17h44m53s.1}, \timeform{-29D31'45''.5}).
Since the flux was almost the same as that of the Suzaku source,
we safely concluded these point sources are identical.
We, then fine-tuned the Suzaku coordinate by shifting
$\Delta$(R.~$\!\!$A., Decl.)=(\timeform{13''.7}, \timeform{11''.2}).

%First, checking the spectra of calibration sources during the observation 
%we found the gain discrepancy of $+4\sim 15{\rm eV}$ @ 5.89~keV 
%among sensors. We corrected the inconsistency before all of the analyses.

\begin{figure}
  \begin{tabular}{cc}
    \hspace{-1cm}
    \begin{minipage}{0.5\textwidth}
%      \FigureFile(91mm,78mm){64clumpregion_245kev_rev6.ps}
      \FigureFile(91mm,78mm){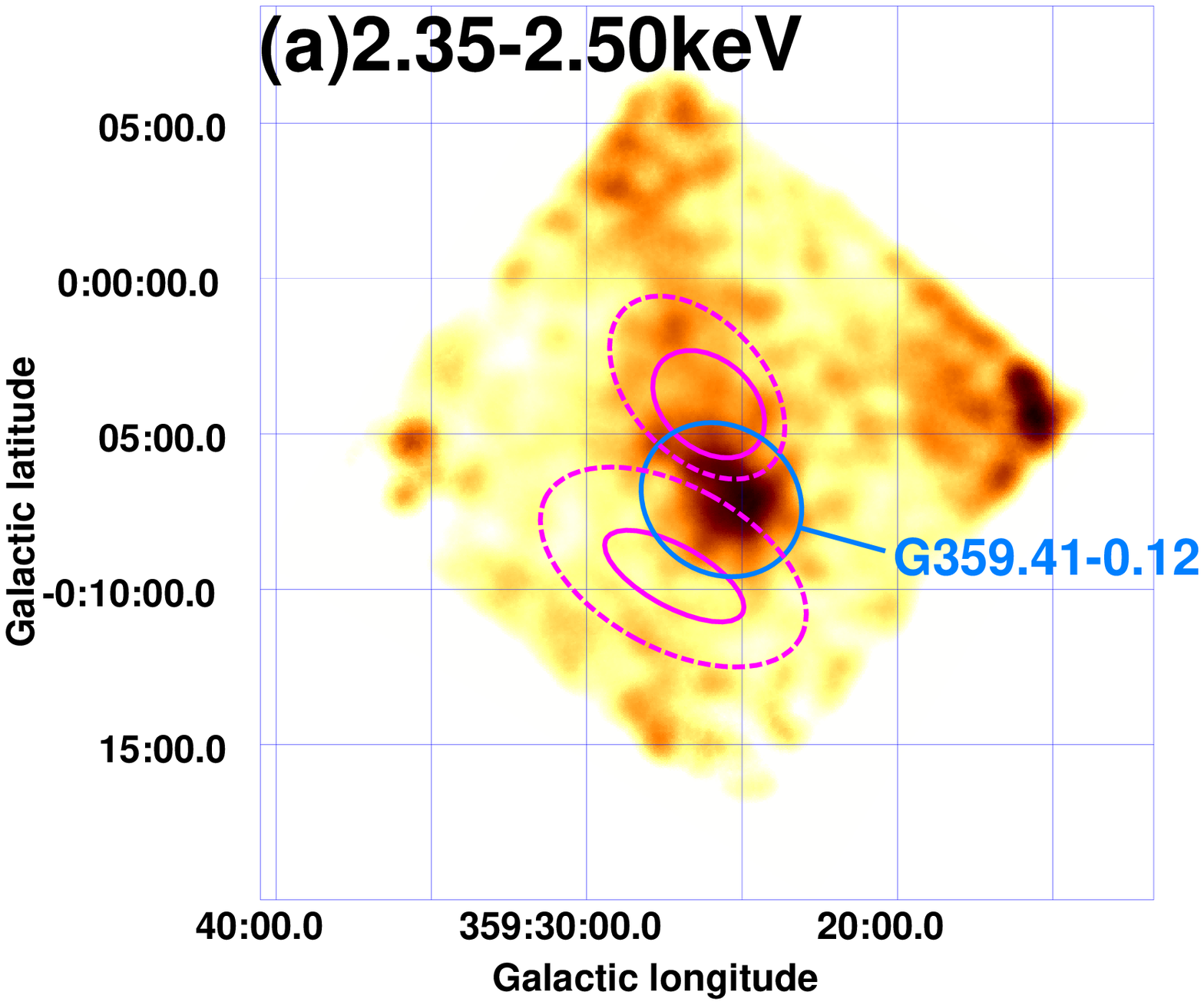}
    \end{minipage}
    \begin{minipage}{0.5\textwidth}
%      \FigureFile(91mm,78mm){64clumpregion_64kev_rev8.ps}
      \FigureFile(91mm,78mm){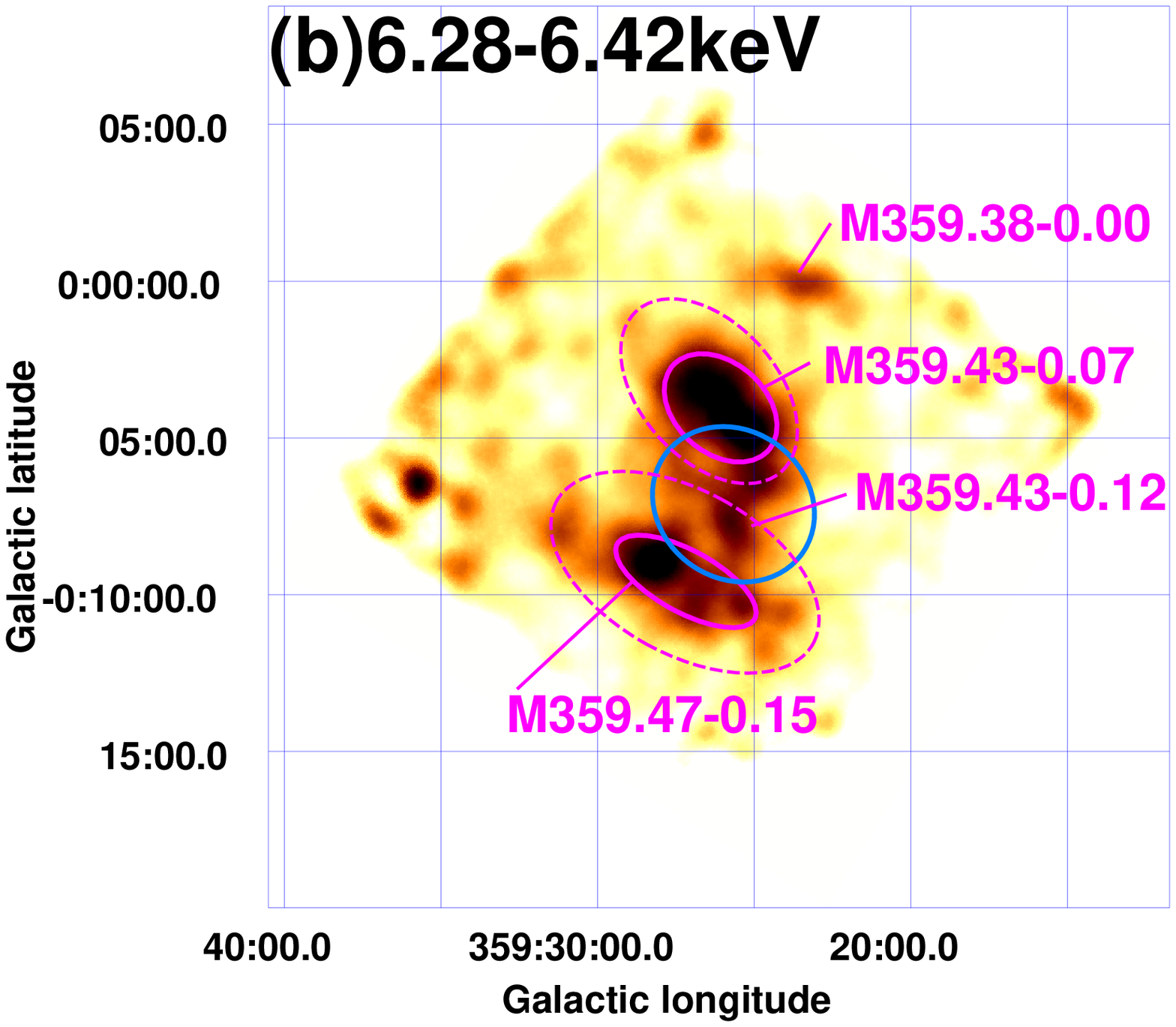}
    \end{minipage}
  \end{tabular}

  \caption{(a) The 2.45~keV-line (K$\alpha$-S\emissiontype{XV}) image
	in the energy band of 2.35--2.50~keV and (b) the 6.4~keV-line 
	(K$\alpha$ of Fe\emissiontype{I}) image in the 6.28--6.42~keV band.
	The source and background regions for the spectral analysis are shown with
	the solid and dashed magenta ellipses, where the data within the blue ellipse
	(G\,359.41$-$0.12) were excluded to minimize the contamination from
	the strong 2.45~keV emission line and its associated continuum.
	}
  \label{fig:suzakuimgs}
\end{figure}

Referring figure~\ref{fig:Spec_circle400_xis023_grp500}, we made
the narrow band image of the 2.45~keV-line (2.35--2.50~keV; 
K$\alpha$ of S\emissiontype{XV}) and the 6.4~keV-line 
(6.28--6.42~keV; K$\alpha$ of Fe\emissiontype{I}).
The images after the smoothing with the Gaussian kernel of 48 pixels
($\timeform{0'.83}$) are shown in figure~\ref{fig:suzakuimgs}.
%%%特に我々が注目する構造について
In figure~\ref{fig:suzakuimgs}(a), a bright 2.45~keV-line clump is found at 
$(l,b)=(\timeform{359D.407}, \timeform{-0D.119})$,
hence designated as G\,359.41$-$0.12.
Likely, four 6.4~keV-line clumps found in figure~\ref{fig:suzakuimgs}(b)
are designated as M\,359.43$-$0.07, M\,359.47$-$0.15, M\,359.43$-$0.12, and 
M\,359.38$-$0.00, where M\,359.43$-$0.12 is in the region of G\,359.41$-$0.12.

\subsection{The Emission Line Feature in the Sgr C Region}\label{ssec:specanaI}

The photon statistics are limited to make measurable spectra for 
M\,359.38$-$0.00 and M\,359.43$-$0.12. 
Furthermore, M\,359.43$-$0.12 is in the 2.45~keV-source, G\,359.41$-$0.12. 
Therefore we concentrated on the two bright sources, 
M\,359.43$-$0.07 and M\,359.47$-$0.15, and the surrounding background regions. 

As is seen in figure~\ref{fig:Spec_circle400_xis023_grp500},
the GCDX with strong 6.7~keV (K$\alpha$ of Fe\emissiontype{XXV})
and 6.9~keV (Ly$\alpha$ of Fe\emissiontype{XXVI}) lines is the
largest background for the local enhancements such as the 6.4~keV-clumps. 
In addition, the flux of the GCDX is not uniform but variable from 
position to position.
To see the flux distribution and nature of the GCDX near Sgr~C,
we examined the NXB-subtracted (GCDX is not subtracted) spectra in the 6.4-keV 
clumps and the surrounding areas including candidate background regions 
(see figure~\ref{fig:suzakuimgs}).

The spectra from the whole Sgr~C region is already made in section~\ref{ssec:image}
(figure~\ref{fig:Spec_circle400_xis023_grp500}). 
The spectra of M\,359.43$-$0.07 and M\,359.47$-$0.15 are extracted from
the regions shown with the solid magenta ellipses in figure~\ref{fig:suzakuimgs}(b).
The background spectra for these 6.4~keV clumps were obtained 
from the annuli between the inner solid and outer dashed ellipses. 
Since G\,359.41$-$0.12, the solid blue ellipse
in figure~\ref{fig:suzakuimgs}(a), may have strong 2.45~keV-line with 
an associated continuum emission, we excluded G\,359.41$-$0.12 from the spectra of
the 6.4-keV sources and their backgrounds.
As the surrounding area, we made additional spectrum from a
circle at the FOV center with a radius of \timeform{7'.8}
but excluding the above cited source and background regions
(hereafter, the outer region). 

\begin{figure}[h]
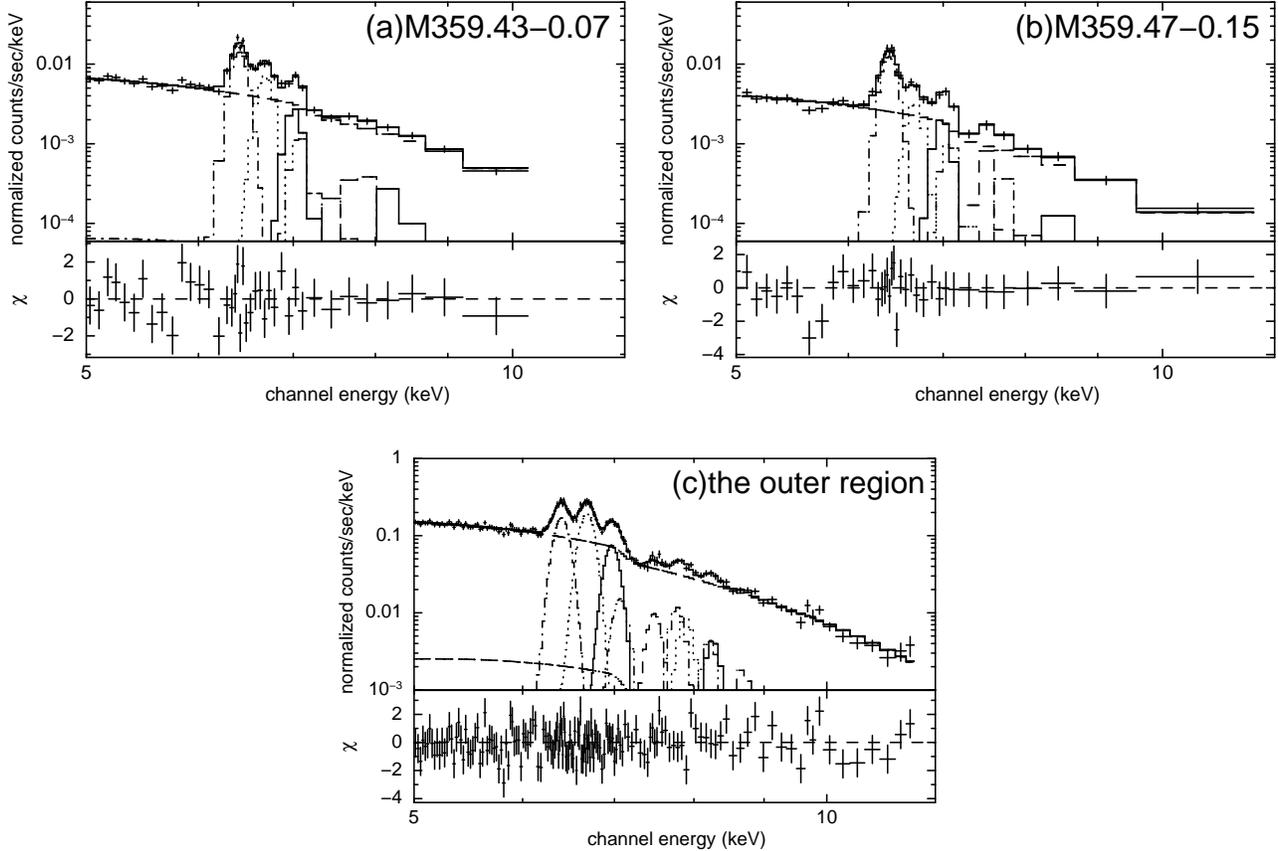

  \begin{tabular}{cc}
    \begin{minipage}{0.5\textwidth}
%      \rotatebox{-90}{\FigureFile(54mm,45mm){2vphabs_pow_onlynxbsub_src3_gaincor_tsuru1_onlyFI.ps}}
      \rotatebox{-90}{\FigureFile(54mm,45mm){figure3a.ps}}
    \end{minipage}
    \begin{minipage}{0.5\textwidth}
%      \rotatebox{-90}{\FigureFile(54mm,45mm){2vphabs_pow_onlynxbsubt_src4_gaincor_tsuru1_onlyFI.ps}}
      \rotatebox{-90}{\FigureFile(54mm,45mm){figure3b.ps}}
    \end{minipage}
    \vspace{0.5cm}\\
    \begin{minipage}{0.5\textwidth}
%      \rotatebox{-90}{\FigureFile(54mm,45mm){ex-clumps_fit_080703_t0_onlyFI.ps}}
      \rotatebox{-90}{\FigureFile(54mm,45mm){figure3c.ps}}
    \end{minipage}
  \end{tabular}
  \caption{Same as figure~\ref{fig:Spec_circle400_xis023_grp500},
	but for M\,359.43$-$0.07 (a),
	M\,359.47$-$0.15 (b), and the outer region (c).
	The phenomenological model with single power-law,
	10 Gaussian lines, and the cosmic X-ray background component is applied.
	}
  \label{fig:spec64kev_onlynxb}
\end{figure}

The spectral fittings were made in the same energy band 
with the same phenomenological model and free parameters 
as those used in Koyama et al.~(2007b, 2008b): 
a single power-law and 10 Gaussian emission lines. 
The spectra of FIs and BI were simultaneously fitted, and
acceptable $\chi^2$ were obtained from all the spectra.
The FI spectra and the best-fit model of M\,359.43$-$0.07 and M\,359.47$-$0.15, 
and the outer region of Sgr~C are shown in figure~\ref{fig:spec64kev_onlynxb} as examples.
The best-fit parameters are listed in table~\ref{tab:fitpara_64kev_onlynxb}.

\begin{table*}[htb]
  \caption{The best-fit parameters for the NXB subtracted spectra.}
  \label{tab:fitpara_64kev_onlynxb}
  \begin{center}
    \begin{footnotesize}
    \begin{tabular}{lcccccc}
      \hline
      Region & Area & $\Gamma$ & \multicolumn{2}{c}{Line Flux$^*$} &
      \multicolumn{2}{c}{Line Equivalent width (eV)} \\
      & (arcmin$^2$) & & $F_{6.4}$ & $F_{6.7}$ & $EW_{6.4}$ & $EW_{6.7}$\\
      \hline
	(1) & 152  & $1.72 (1.65-1.78)$ & $0.79 (0.77-0.81)$ & $0.68 (0.65-0.70)$
	& $458 (448-470)$ & $420 (401-430)$\\
	(2) & 121  & $1.70 (1.63-1.82)$ & $0.55 (0.53-0.57)$ & $0.68 (0.66-0.71)$
	& $346 (333-363)$ & $454 (441-471)$\\
	(3) & 7.09 & $1.82 (1.65-2.06)$ & $1.91 (1.79-2.06)$ & $1.08 (0.96-1.22)$
	& $670 (625-720)$ & $400 (357-454)$\\
	(4) & 11.7 & $1.64 (1.47-1.93)$ & $1.00 (0.92-1.08)$ & $0.93 (0.86-1.03)$
	& $397 (365-428)$ & $388 (360-431)$\\
	(5) & 6.91 & $1.96 (1.59-2.24)$ & $2.16 (2.03-2.31)$ & $0.63 (0.53-0.73)$
	& $966 (909-1030)$ & $298 (253-348)$\\
	(6) & 22.4 & $1.75 (1.59-1.97)$ & $1.02 (0.97-1.08)$ & $0.67 (0.62-0.72)$
	& $587 (554-620)$ & $404 (374-437)$\\
      \hline 
      \multicolumn{7}{@{}l@{}}{\hbox to 0pt{\parbox{130mm}{\footnotesize
      (1) the whole Sgr~C region, (2) the outer region, (3) M\,359.43$-$0.07,
      (4) background region for M\,359.43$-$0.07, (5) M\,359.47$-$0.15, 
      (6) background region for M\,359.47$-$0.15.
      \par\noindent
      90\% confidence limits are in parentheses. 
      \par\noindent
      \footnotemark[$*$] 
      Absorption-uncorrected line flux in the unit of 
      $10^{-6}$ph\ cm$^{-2}$\ s$^{-1}$\ arcmin$^{-2}$.
      }\hss}}
    \end{tabular}
    \end{footnotesize}
  \end{center}
\end{table*}

\citet{Koyama08b} found that the GCDX in the Sgr~A region
is phenomenologically decomposed into the 6.7-keV line plus
associated continuum (6.7-component) and the 6.4-keV line
plus associated continuum (6.4-component);
the EWs of the 6.4~keV ($EW_{6.4}$) and the 6.7~keV ($EW_{6.7}$)
lines are given by the relation
$EW_{6.7} + 0.50(\pm0.06) \times EW_{6.4} = 0.62(\pm0.07) {\rm [keV]}$.
Then, in the limit of $EW_{6.4}\to 0$, $EW_{6.7}$ was estimated to be
$0.62 \pm 0.07$~keV.
On the other hand, in the limit of $EW_{6.7}\to 0$, $EW_{6.4}$ is $1.2\pm 0.2$~keV.
They also made the GCDX-subtracted spectra of the two 6.4~keV clumps in
the Sgr~A region (source 1 and 2 in \cite{Koyama08b})
and obtained  consistent $EW_{6.4}$ values as those estimated from the
phenomenological relation.

\begin{figure}
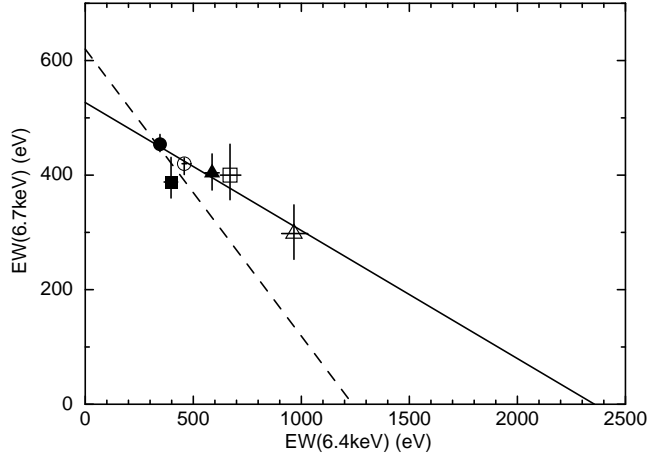

    \begin{center}
%      \rotatebox{-90}{\FigureFile(60mm,30mm){EWEWralation_rev1.ps}}
      \rotatebox{-90}{\FigureFile(60mm,30mm){figure4.ps}}
    \end{center}
      \caption{The correlation between the EWs of the 
        Fe\emissiontype{I}-K$\alpha$ line ($EW_{6.4}$) and 
        Fe\emissiontype{XXV}-K$\alpha$ line ($EW_{6.7}$).
        The dashed line shows the best fit relation in the Sgr~A region 
        (adopted from figure~2(d) of \cite{Koyama08b}). 
        The open and filled circle show the whole Sgr~C region 
        and the outer region, respectively. 
        The source regions of M\,359.43$-$0.07 and M\,359.47$-$0.15 
        are plotted with the open square and triangle, while their background 
        regions are given by the filled marks. 
        The solid line is the best-fit relation for all regions except the whole
        Sgr~C of $EW_{6.7} + 0.22 \times EW_{6.4} = 0.53 {\rm [keV]}$. 
       }
  \label{fig:EWcorr}
\end{figure}

Stimulated by the successful approach of \citet{Koyama08b},
we used the same method to decompose the GCDX in the Sgr~C region
into the 6.4-component and 6.7-component.
Using the data in table~\ref{tab:fitpara_64kev_onlynxb}, we plot
the $EW_{6.4}$ and $EW_{6.7}$
relation of the six regions (figure~\ref{fig:EWcorr}). 
The outer region, the whole Sgr~C and the background region 
for M\,359.43$-$0.07 follow to, but the M\,359.43$-$0.07 and M\,359.47$-$0.15
regions come above the best-fit correlation line of the Sgr~A region;
these are systematically larger $EW_{6.4}$ than the Sgr~A region. 
The best-fit linear relations between $EW_{6.7}$ and $EW_{6.4}$ for all cited
regions except the whole Sgr~C is; 

\begin{equation}
  EW_{6.7} + 0.22(\pm0.12) \times EW_{6.4} 
    = 0.53(\pm0.06) {\rm [keV]}, 
\end{equation}
where the errors mean 90\% confidence levels.

This relation indicates that $EW_{6.7}$ in the 6.7-component 
is $0.53 \pm 0.06$~keV (at $EW_{6.4}\to 0$)
for the Sgr~C region, consistent with that of the Sgr~A region of 
$0.62 \pm 0.07$~keV (\cite{Koyama08b}). 
On the other hand, $EW_{6.4}$ in the 6.4-component is 
$2.4^{+2.3}_{-0.7}$~keV (at $EW_{6.7}\to 0$), 
larger than those in the Sgr~A region \citep{Koyama08b}. 

\subsection{Spectra of M\,359.43$-$0.07 and M\,359.47$-$0.15}\label{ssec:specanaII}

From table~\ref{tab:fitpara_64kev_onlynxb}, we see that the 6.7~keV line flux 
in M\,359.43$-$0.07 and M\,359.47$-$0.15 are almost identical to those of
the respective background regions. The 6.7~keV flux differences are 
16\% and 6\% of those in M\,359.43$-$0.07 and M\,359.47$-$0.15, respectively. 
Since the 6.7~keV line flux is a good indicator of the flux of the 
GCDX (Koyama et al. 2007b, 2008b), we conclude
that the background regions were properly selected within
possible systematic error of 6--16\%, which is smaller than the statistical errors.

\begin{figure}
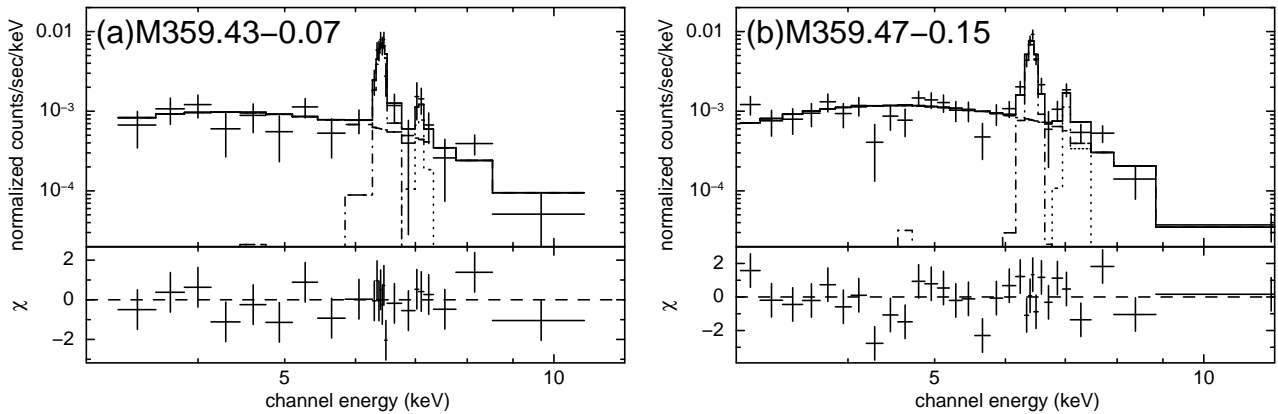

  \begin{tabular}{cc}
    \begin{minipage}{0.5\textwidth}
%      \rotatebox{-90}{\FigureFile(54mm,45mm){2vphabs_pow_bgdnorm_src3bgd6_incBI_from3keV_gaincor_n1.ps}}
      \rotatebox{-90}{\FigureFile(54mm,45mm){figure5a.ps}}
    \end{minipage}
    \begin{minipage}{0.5\textwidth}
%      \rotatebox{-90}{\FigureFile(54mm,45mm){2vphabs_pow_bgdnorm_src4bgd4_incBI_from3keV_gaincor.ps}}
      \rotatebox{-90}{\FigureFile(54mm,45mm){figure5b.ps}}
    \end{minipage}
  \end{tabular}
  \caption{Same as figure~\ref{fig:spec64kev_onlynxb}, but for GCDX-subtracted spectra of 
	M\,359.43$-$0.07 (a) and M\,359.47$-$0.15 (b).
	The K$\alpha$ and K$\beta$ emission lines of 
    Fe\emissiontype{I} and power-law are shown by dot-dashed,
    dotted, and dashed lines, respectively. 
  }
  \label{fig:spec64kev}
\end{figure}

Therefore we made the GCDX-subtracted spectra of these 
6.4~keV clumps (figure~\ref{fig:spec64kev}).
The energy-dependent vignetting were corrected by multiplying the effective-area
ratio of the source region to corresponding background region for each
energy bin of the GCDX spectra.

We first applied a model of a power-law and a narrow Gaussian line at 6.4~keV
with absorption. The metal abundances of the inter stellar 
medium were fixed to the solar value \citep{Anders89}, 
while the center energy of the 6.4~keV line and its flux were free parameters. 
We simultaneously fit this model to the FIs and BI spectra,
and found a line-like residual at 7.0--7.1~keV,
with the flux of $\sim$10--20\% of the 6.4~keV (Fe\emissiontype{I} K$\alpha$) line.
From the energy and flux, the residual is likely to be the 
K$\beta$ of neutral Fe. We then added second narrow Gaussian for 
the Fe\emissiontype{I} K$\beta$ line, fixing its center energy to
1.103 times of that of K$\alpha$.
This model was accepted as is shown in figure~\ref{fig:spec64kev},
while the best-fit parameters are listed in table~\ref{tab:fitpara_64kev}. 

\begin{table*}[htb]
  \caption{The best-fit parameters of the spectral fittings to the 
    background (GCDX) subtracted spectra of M\,359.43$-$0.07 and M\,359.47$-$0.15. }
  \label{tab:fitpara_64kev}
  \begin{center}
    \begin{tabular}{lcc}
      \hline
      Parameter & M\,359.43$-$0.07 & M\,359.47$-$0.15 \\
      \hline
      Absorbed power-law model: & & \\
      \ \ Column density $N_{\rm H}$ ($10^{23}{\rm cm}^{-2}$) & 
      $0.92 (0.48 - 1.41)$ &
      $0.82 (0.65 - 1.18)$ \\
      \ \ Photon index &
      $1.67 (1.51 - 1.82)$ &
      $1.61 (1.52 - 1.86)$ \\
      Gaussian 1 (FeI K$\alpha$): & & \\
      \ \ Line energy (keV) &
      $6.41 (6.39 - 6.42)$ &
      $6.41 (6.40 - 6.42)$ \\
      \ \ Line flux (10$^{-6}$~ph~cm$^{-2}$~s$^{-1}$) $^*$&
      $6.43 (5.27 - 7.39)$ &
      $8.82 (7.87 - 10.0)$ \\
      \ \ Equivalent width (keV) &
      $2.18 (1.79 - 2.51)$ &
      $1.96 (1.75 - 2.23)$ \\
      Gaussian 2 (FeI K$\beta$): & & \\
      \ \ Line energy (keV)$^\dagger$ &
      $7.07$  &
      $7.07$ \\
      \ \ Line flux (10$^{-6}$~ph~cm$^{-2}$~s$^{-1}$) $^*$&
      $1.05 (0.43 - 2.14)$ &
      $1.91 (1.22 - 3.00)$ \\
      \hline
      Flux (10$^{-13}$erg~cm$^{-2}$~s$^{-1}$)$^\ddagger$ &
      2.70 &
      4.11 \\
      $\chi^{2}$/d.o.f. &
      $27.92/41$ &
      $49.67/46$ \\
      \hline
      \multicolumn{3}{@{}l@{}}{\hbox to 0pt{\parbox{150mm}{\footnotesize
%           \footnotemark[$*$] 
%            この表は
%            (M359.43--0.07)
%            2vphabs_pow_bgdnorm_src3bgd6_incBI_from3keV_gaincor_t1.log
%            (M359.47--0.15)
%            2vphabs_pow_bgdnorm_src4bgd4_incBI_from3keV_gaincor_t1.log
%            から得た。
        90\% confidence limits are in parentheses.
	    
	    \footnotemark[$*$] 
            Line fluxes are not corrected for the absorption.

        \footnotemark[$\dagger$] 
	    The line center is fixed so that the energy ratio is 1.103
	    (K$\beta$/K$\alpha$=$7.058{\rm keV}/6.400{\rm keV}$). 

	    \footnotemark[$\ddagger$]
            Observed flux of the diffuse emission 
	    in the energy band of 3 -- 10~keV. 
            Absorption is not corrected. 
	  }\hss}}
    \end{tabular}
  \end{center}
\end{table*}

%%%%%%%%%%%%%%%%%%%%%%%%%%%%%%%%%%%%%
% Discussion
%
\section{Discussion}

We detected four 6.4~keV clumps, M\,359.43$-$0.07, M\,359.47$-$0.15,
M\,359.43$-$0.12, and M\,359.38$-$0.00 with Suzaku.
From the former bright clumps, M\,359.43$-$0.07 and M\,359.47$-$0.15,
we found K$\alpha$ and K$\beta$-lines from Fe\emissiontype{I}.  
The most important fact is that EWs of the K$\alpha$-lines are 
extremely large as $\sim$2~keV (section~\ref{ssec:specanaII}).
These large EWs are independently 
supported by the analysis of the GCDX in the Sgr C regions
(section~\ref{ssec:specanaI}), and hence are highly reliable.
We note that \citet{Yusef07} reported small $EW_{6.4}$ from all the clumps,
but their estimation was based on the spectra where no GCDX is subtracted.

\subsection{Comparison with the Past X-Ray Observations}
\label{ssec:pastXrayobs}

\citet{Murakami01b} found excess emission at 
$(l,b)=(\timeform{359D.43}, \timeform{-0D.04})$ in the X-ray image of
the 5.8--7.0~keV band. Although the energy resolution was limited, 
an emission line at the  energy consistent with Fe\emissiontype{I}-K$\alpha$ line
was found.  Hence \citet{Murakami01b} regarded this clump as an XRN.

The Suzaku source M\,359.43$-$0.07 is located in the ASCA clump,
but the peak position is systematically shifted. For comparison,
we extracted the Suzaku spectrum from the same source and background regions
as those of the ASCA clump \citep{Murakami01b}. 
The background subtracted spectrum has the 6.4~keV line flux 
of $1.8^{+0.3}_{-0.2}\times 10^{-5}$~ph~cm$^{-2}$~s$^{-1}$ (absorption uncorrected), 
which agrees with that of the ASCA clump
($3.5^{+1.4}_{-2.2}\times 10^{-5}$~ph~cm$^{-2}$~s$^{-1}$).
With Suzaku, we found that the surface brightness ratio of the 6.4~keV 
emission lines between the M\,359.43$-$0.07, the ASCA clump and the outer 
background region are approximately 3:2:1
(1.9, 1.4, and 0.6 in the unit of $10^{-6}$~ph~cm$^{-2}$~s$^{-1}$~arcmin$^{-2}$).
Hence no significant detection of
the full region of the ASCA clump in figure~\ref{fig:suzakuimgs}(b)
may be due to its relatively lower surface brightness 
than M\,359.43$-$0.07. On the other hand, no ASCA detection of M\,359.43$-$0.07
is puzzling, although possible time variability can not be excluded.
ASCA also detected no flux from the other bright Suzaku source M\,359.47$-$0.15.
However, this source is near the edge of the ASCA field.

The Suzaku spectrum of the whole Sgr~C region (figure~\ref{fig:Spec_circle400_xis023_grp500}) 
shows the EWs of $458^{+12}_{-10}$~eV and $420^{+10}_{-19}$~eV 
for the 6.4~keV and 6.7~keV emission lines, respectively. 
The surface brightness in the 5--8~keV energy band is
$7.1\times 10^{-14}$~erg~cm$^{-2}$~s$^{-1}$~arcmin$^{-2}$.
The Chandra analysis of the same region, in which point sources cataloged 
in the list of \citet{Muno06} are excluded, shows that the EWs of the 6.4~keV 
and 6.7~keV emission lines are $460\pm 100$~eV and $\sim 400$~eV, respectively. 
The surface brightness in the 5--8~keV  band is 
$\sim7.5\times 10^{-14}$~erg~cm$^{-2}$~s$^{-1}$~arcmin$^{-2}$ (figure~1b in \cite{Yusef07}). 
These are very similar to those of Suzaku, and hence point source contribution
with the flux level larger than  $1\times 10^{-13}$erg~cm$^{-2}$~s$^{-1}$ \citep{Muno06} 
may not be large, at least in the whole Sgr C region.

Detailed comparison in the smaller scale emission, however, shows significant differences.
M\,359.47$-$0.15 and a part of M\,359.43$-$0.07 can be seen in the narrow band image
at 6.4~keV with Chandra (figure~5a in \cite{Yusef07}).
However, M\,359.43$-$0.07 is not found in the EW
map of Chandra (figure~5c in \cite{Yusef07}). This is a puzzle, because the Suzaku
EW of M\,359.43$-$0.07 is one of the largest among cited regions in the NXB-subtracted spectra.
One possibility is a time variability as was found in Sgr~B2 and Radio Arc regions
\citep{Muno07, Koyama08a, Inui08}.
In the Chandra image of the 2--6~keV band (figure~1a of \cite{Yusef07}),
a clump named as G\,359.45$-$0.07 is found. 
However Suzaku found no excess in this region with the 6.4 keV line, 
and hence G\,359.45$-$0.07 would not be 6.4~keV line emitter.
Conversely no excess from M\,359.38$-$0.00
is found with either Chandra or ASCA, possibly due to limited flux, and hence
this is a new 6.4~keV clump found with the Suzaku satellite.

\subsection{Ionization Mechanism}

Two scenarios have been proposed for the origin of the 6.4~keV 
emission line from MCs in the GC region. One is X-ray photo ionization 
by external X-ray sources (the XRN scenario) \citep{%Sunyaev93, 
Koyama96, Sunyaev98, Park04}. 
The other is the inner shell ionization by the impact of LECRe
(the electron bombardment scenario) \citep{Valinia00, Yusef02}.
The LECRe scenario 
expects  $EW_{6.4}$ of $\sim 300$ eV for the solar abundance Fe
\citep{Tatischeff03, Yusef07}, and hence is difficult to explain the observed large 
$EW_{6.4}$ of $\sim 2$ keV, unless Fe is extremely over-abundant by a 
factor of 6--7. We found no observational evidence for such over-abundance.
On the other hand, XRN scenario naturally explain the
large $EW_{6.4}$ of the M\,359.43$-$0.07 and M\,359.47$-$0.15,
and hence is more likely.

\subsection{Photo-Ionization Source}

Since the typical absorption toward GC is
$\sim 0.6\times 10^{23}$~H~cm$^{-2}$ \citep{Rieke89,Sakano02},
intrinsic absorption depth of M\,359.43$-$0.07 and M\,359.47$-$0.15 are
$\lesssim 0.8\times10^{23}$~H~cm$^{-2}$ and $\lesssim 0.6\times 10^{23}$~H~cm$^{-2}$,
respectively. The areas of M\,359.43$-$0.07 and M\,359.47$-$0.15 shown in
figure~\ref{fig:suzakuimgs}(b) are 7~${\rm arcmin^2}$, 
then their masses are estimated to be $\lesssim 4\times10^4$, 
$\lesssim 3\times 10^4$ solar mass, respectively. 
From these masses and fluxes of the 6.4~keV lines, the luminosity of 
photo-ionizing source can be estimated following \citet{Sunyaev98}
as $\gtrsim1\times10^{38}$~erg~s$^{-1}$  
and $\gtrsim2\times10^{38}$~erg~s$^{-1}$.

There is no X-ray object bright enough inside or nearby Sgr~C region. 
Even the brightest near-by sources, KS\,1741$-$293 and the Great Annihilator
(1E\,1740.7--2943)
are impossible to explain this luminosity. 
Thus we arrive to the same conclusion for the 6.4~keV clumps origin near Sgr B2,
a past Sgr~A$^*$ activity of $\gtrsim1\times10^{38}$~erg~s$^{-1}$  at 240~yr 
ago.

\subsection{Comparison with the Molecular Line Observations}\label{ssec:comparisonradio}

\begin{figure}
\begin{center}
% \FigureFile(105mm,90mm){64clumpregion_64kev_rev9_cscontours3080_90130.ps}
 \FigureFile(105mm,90mm){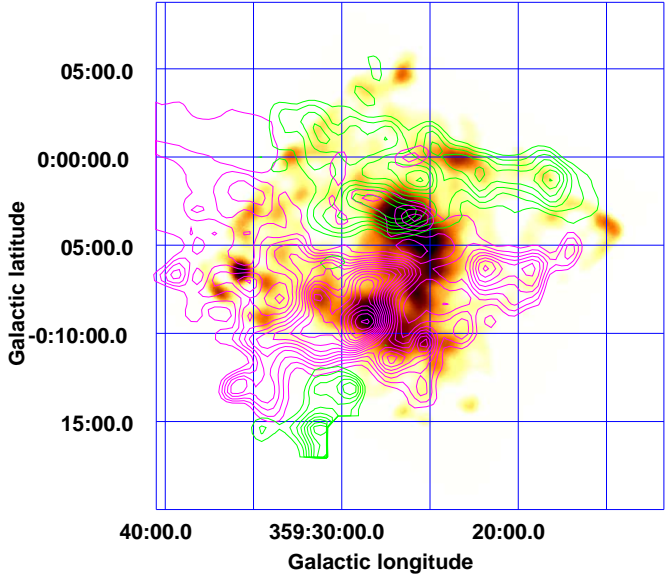}
  \end{center}
  \caption{
    The contours of channel map of CS $J = 1-0$ emission line 
    overlaid on the narrow band image at 6.4~keV shown 
    in figure \ref{fig:suzakuimgs}(b). 
    The first contour level and contour interval are 
    14.3 and 4.8~K~km~s$^{-1}$, respectively.
    Emission in the velocity ranges of $-130$ to $-90$~km s$^{-1}$ is shown
    with green and that of $-80$ to $-30$~km s$^{-1}$ is magenta,
    respectively.}
  \label{fig:xray-cs}
\end{figure}

The absorption column derived from the background-subtracted spectra
of M\,359.43$-$0.07 and M\,359.47$-$0.15 are
$N_{\rm H}$$\sim 1\times10^{23}$~H~cm$^{-2}$.
We require similar amounts of column in the MCs
that corresponds to M\,359.38$-$0.00 and M\,359.43$-$0.12. 
Referring the CS molecular line map over the velocity range of $V_{\rm LSR}= -200$
to +200~km~s$^{-1}$ (\cite{Tsuboi99}, and the electric data),
we searched for possible counterpart of the 6.4~keV clumps with 
$N_{\rm H}$$\geq10^{23}$~H~cm$^{-2}$ (figure~\ref{fig:xray-cs}). 

M\,359.43$-$0.07 :  MCs of $N_{\rm H}$$\sim 10^{23}$H~cm$^{-2}$ 
are seen both in the velocity ranges of $-50$ to $-70$~km~s$^{-1}$ 
(also \cite{Yusef07}) and $-90$ to $-110$~km~s$^{-1}$ (also \cite{Murakami01b}).

M\,359.47$-$0.15 : An MC with $N_{\rm H}$$\ge 10^{23}$~H~cm$^{-2}$ 
in the velocity range of $-50$ to $-70$~km~s$^{-1}$ exists in this region. 
M\,359.47$-$0.15 is located in the western edge of $-65$~km~s$^{-1}$ MC
as is also noticed by \citet{Yusef07}. There is no other MC of $\sim 10^{23}$~H~cm$^{-2}$
in the other velocity range. 

M\,359.38$-$0.00 : No MC is seen in the $-50$ to $-70$~km~s$^{-1}$ band
but an MC in the velocity range of $-110$ to 
$-130$~km~s$^{-1}$ is associated to the 6.4-keV clump; 
The column density of this radio cloud is roughly half of those in the directions 
of M\,359.43$-$0.07 and M\,359.47$-$0.15.

M\,359.43$-$0.12 : 
There is a weaker peak in the $-130$ to $-120$~km~s$^{-1}$ band than
that at M\,359.38$-$0.00. 

\citet{Sofue95} and \citet{Sawada04} proposed that molecular gases in the GC region 
consists of two arms, Arm~I and Arm~II. 
The $-130$ to $-90$~km~s$^{-1}$ MC is physically
associated with Arm~I, while the $-80$ to $-30$~km~s$^{-1}$ MC
is a part of Arm~II. 
M\,359.38$-$0.00 and M\,359.47$-$0.15 belong to Arm~I and Arm~II, respectively,
while M\,359.43$-$0.07 has both possibilities. 
Since Arm~II is located far side of Arm~I \citep{Sofue95, Sawada04}, 
X-rays from Sgr~A$^*$ of several hundred years ago arrived
at M\,359.47$-$0.15 probably earlier than M\,359.38$-$0.00. 

%\subsection{Time Variable X-rays from Sgr~A$^*$}
%\citet{Yusef07} argued against the XRN model for the Sgr~C region
%from two pieces of observational evidence; 
%(1) M\,359.47$-$0.15 is located in the far side of the $-110$~km~s$^{-1}$ MC
% from Sgr~A$^*$, while near side of the MC is faint in 6.4~keV,  
%and (2) only a small fractions of the MCs 
%from $-200$~km~s$^{-1}$ to +200~km~s$^{-1}$ emit measurable  6.4~keV line. 
%\citet{Muno07}, \citet{Koyama08a}, and \citet{Inui08} discovered the time variability of 
%the 6.4~keV clouds in the Sgr~A and Sgr~B2 region. 
%In the XRN scenario, these observational results indicate 
%that the luminosity of  Sgr~A$^*$ was time variable 
%with the timescales of $2 \sim 10$~yr. 
%The typical projected size of the 6.4~keV clumps in Sgr~C region 
%is $10\sim 20$~ly, which matches the flare timescale. 
%Some differences in  small scale ($\sim$10~pc) 6.4 keV structures
%between the ASCA, Chandra and Suzaku observations as reported in section 4.1 
%may be due to this possible time variability. To confirm any time variability
%in the Sgr~C region, however, we need follow-up deep exposure observations. 

\section{Summary}\label{sect:summary}
\begin{enumerate}

\item We found four diffuse 6.4~keV clumps, M\,359.43$-$0.07, M\,359.47$-$0.15,
M\,359.43$-$0.12 and M\,359.38$-$0.00 in the Sgr~C region. The last one, 
M\,359.38$-$0.00 is newly discovered. 

\item  The spectra of the two bright clumps, M\,359.43$-$0.07 and
  M\,359.47$-$0.15, have power-law of photon index of 1.6--1.7 with
  a very large $EW_{6.4}$ of 2.0--2.2~keV, the largest $EW_{6.4}$
  among the 6.4~keV clumps in the GC.

\item The large $EW_{6.4}$ supports that the origin of 6.4~keV clumps 
is due to X-ray reflection irradiated by hard X-ray sources.

\item M\,359.38$-$0.00 and M\,359.47$-$0.15 possibly associate with MCs in 
different velocity ranges, indicating these are located in the  different 
molecular arms.

\end{enumerate}
\bigskip
We are grateful to all the members of the Suzaku hardware and software
teams, and the science working group.
This work was supported by the Grant-in-Aid for the Global COE Program
``The Next Generation of Physics, Spun from Universality and Emergence'' 
from the Ministry of Education, Culture, Sports, Science and Technology (MEXT) of Japan.
HN and YH are financially supported by the Japan Society for the Promotion of Science.

%%%%%%%%%%%%%%%%%%%%%%%%%%%%%%%%%%%%%%%
%%%
% See the manual for the detail.
%%%

\end{document}